\documentclass[fleqn,12pt]{wlscirep}
\usepackage[utf8]{inputenc}
\usepackage[utf8]{inputenc}
\usepackage{braket}
\usepackage{algorithm,algcompatible,amsmath}
\usepackage{subfigure}
\usepackage{comment}
\usepackage{url}
\algnewcommand\INPUT{\item[\textbf{Input:}]}%
\algnewcommand\OUTPUT{\item[\textbf{Output:}]}%

\usepackage[super,sort&compress,comma]{natbib}
\usepackage{graphicx}

\usepackage{xcolor}
\usepackage[normalem]{ulem}

\newcommand{\todo}[1]{\textcolor{red}{}}

\usepackage{natbib}
\usepackage{graphicx}
\usepackage[T1]{fontenc}
\title{Variational quantum eigensolver for closed-shell molecules with non-bosonic corrections}

\author[1,$\dagger$]{Kyungmin Kim}
\author[2,$\dagger$]{Sumin Lim}
\author[3,*]{Kyujin Shin}
\author[4]{Gwonhak Lee}
\author[5]{Yousung Jung}
\author[3]{Woomin Kyoung}
\author[4,6,7,*]{June-Koo Kevin Rhee}
\author[1,*]{Young Min Rhee}

\affil[1]{Department of Chemistry, KAIST, Daejeon, 34141, Republic of Korea}
\affil[2]{Department of Physics, KAIST, Daejeon, 34141, Republic of Korea}
\affil[3]{Materials Research \& Engineering Center, CTO Division, Hyundai Motor Company, Uiwang 16082, Republic of Korea}
\affil[4]{School of Electrical Engineering, KAIST, Daejeon, 34141, Republic of Korea}
\affil[5]{Department of Chemical Engineering, Seoul National University, Seoul, 08826, Republic of Korea}
\affil[6]{KAIST ITRC of Quantum Computing for AI, KAIST, Daejeon, 34141, Republic of Korea}
\affil[7]{KAIST Institute for IT Convergence, KAIST, Daejeon, 34141, Republic of Korea}

\affil[*]{Corresponding authors: shinkj@hyundai.com, rhee.jk@kaist.ac.kr, and ymrhee@kaist.ac.kr}
\affil[$\dagger$]{These authors contributed equally}

\begin{abstract}
The realization of quantum advantage with noisy-intermediate-scale quantum (NISQ) machines has become one of the major challenges in computational sciences. Maintaining coherence of a physical system with more than ten qubits is a critical challenge that motivates research on compact system representations to reduce algorithm complexity. Toward this end, quantum simulations based on the variational quantum eigensolver (VQE) is considered to be one of the most promising algorithms for quantum chemistry in the NISQ era. We investigate reduced mapping of one spatial orbital to a single qubit to analyze the ground state energy in a way that the Pauli operators of qubits are mapped to the creation/annihilation of singlet pairs of electrons. To include the effect of non-bosonic (or non-paired) excitations, we introduce a simple correction scheme in the electron correlation model approximated by the geometrical mean of the bosonic (or paired) terms. Employing it in a VQE algorithm, we assess ground state energies of H$_2$O, N$_2$, and Li$_2$O in good agreements with full configuration interaction (FCI) models respectively, using only 6, 8, and 12 qubits with quantum gate depths proportional to the squares of the qubit counts. With the adopted seniority-zero approximation that uses only one half of the qubit counts of a conventional VQE algorithm, we find our non-bosonic correction method reaches reliable quantum chemistry simulations at least for the tested systems.
\end{abstract}
\begin{document}

\flushbottom
\maketitle
%
%
\thispagestyle{empty}


\section{Introduction}

The concept of quantum simulation using a quantum computer was first proposed by Feynman,\cite{feynman1982simulating} from an insight that a coupled quantum state has the ability to efficiently and accurately simulate another quantum mechanical system. More than a decade later, the conjectured efficiency was confirmed by Lloyd.\cite{lloyd1996universal} In the early developments, the phase estimation algorithm (PEA) proposed by Kitaev was adopted crucially,\cite{kitaev1995quantum} with experimental verification in a small qubit system such as a nuclear magnetic resonance (NMR) device.\cite{du2010nmr} Among the many different possibilities of using quantum algorithms in solving real-world problems, efficiently solving electronic structure problems as suggested by Aspuru-Guzik et al.\cite{aspuru2005simulated}
has drawn much attention. While quantum computers in the future are expected to outperform classic computers for some specific problems, in the viewpoint of scalability we currently only have few tens of noisy qubits as Preskill pointed out.\cite{preskill2018quantum} The present devices are conventionally characterized as noisy-intermediate-scale quantum (NISQ) devices. 
Thus, demonstrating quantum superiority within this limitation is one of the major challenges. 

With this situation, the variational quantum eigensolver (VQE)\cite{peruzzo2014variational} has become likely the primary algorithm for performing quantum chemistry simulations. In the realm of the electronic structure theory, small molecules approximately having 10 electrons are simulated by quantum hardware,\cite{kandala2017hardware, li2019quantum, nam2020ground} while systems that require 20 -- 30 qubits have been calculated on models.\cite{Rice2021JCP, mccaskey2019quantum} Development of VQE algorithms to perform more efficient quantum simulations using NISQ hardware is being reported continuously.\cite{google2020hartree, foxen2020demonstrating}

In electronic structure problems, obtaining the exact solution of the time-independent Schr\"{o}dinger equation, i.e., full configuration interaction (FCI) energy, has a complexity close to $O(N!)$ or exponential for practical purposes with the basis set size $N$. FCI calculation considers all possible electron configurations within the available orbital space beyond the Hartree-Fock (HF) determinant. A VQE approach initiates the post-HF calculation by mapping the $N$ selected spin-orbitals to qubits and generates an ansatz that can be prepared by a system of unitary coupled cluster (UCC) gates or some other heuristic gates. This is followed by the measuring energy expectation value for the corresponding Pauli operators in the computational basis. VQE obtains an approximate value of the FCI energy with a polynomial complexity with respect to $N$, which is attained by adjusting the parameters of ansatz using the classical optimizer. Practically, however, as the numbers of qubits and gates of a noisy quantum computer increase, the fidelity starts to drop rapidly and the VQE algorithm do not achieve the desired accuracy. Therefore, in the NISQ era, constructing an efficient VQE algorithm that reduces the numbers of qubits and gates is one of the most significant approaches. In conventional VQE algorithms, a spin orbital is encoded by a single qubit by Jordan-Wigner\cite{jordan1928pauli} or Bravyi-Kitaev\cite{bravyi2002fermionic} transformation. For the closed-shell molecules, however, a seniority-zero approximation is routinely applied to reduce the total number of qubits required,\cite{elfving2020simulating} which truncates the un-paired excitations in the VQE ansatz. An example of this approximation is doubly occupied configuration interaction (DOCI), where one includes all determinants only with doubly occupied orbitals. Because such pair-correlated methods can capture a significant portion of static correlations, previous studies\cite{stein2014seniority, van2015polynomial} focused on improving the missing dynamic correlations. From the viewpoint of the NISQ device, pair-correlated approximation is promising since the number of qubits required to implement the ansatz can be reduced by a factor of two, because one qubit encodes a spatial orbital,\cite{elfving2020simulating} not a spin-orbital. Indeed, some of the authors have recently demonstrated this advantage with a trapped-ion quantum hardware with the orbital-optimized pair-correlated unitary pair coupled cluster double ansatz (oo-upCCD).\cite{zhao2023} Thus, designing a VQE algorithm that can recover the missing correlation energy of the pair-approximation will be important for the utility of a NISQ quantum computer.

Here, we propose a simple correction scheme for the orbital-optimized pair-correlated VQE simulation. Specifically, we first construct an ansatz by using exchange gates between the qubits corresponding to occupied and virtual spatial orbitals. The essence of this construct is the same as in the earlier studies listed in the above. The VQE optimization within the ansatz and then subsequent measurements provide information needed for further performing orbital optimizations,\cite{stein2014seniority,zhao2023} which we then perform with a classical algorithm.\todo{cite}
For handling electron correlations involving singly occupied orbitals besides the double excitations, we propose a simple non-bosonic correction based on the terms designed with the geometric means of the related bosonic excitation terms. The correction can be performed without using any quantum resource. We test the performance of our scheme by considering a series of molecular systems. Indeed, reasonable agreements with the FCI results are attained for all the tested systems, and the non-bosonic corrections in many cases are shown crucial in achieving the agreements. In fact, the non-bosonic correction scheme that we are proposing is computationally almost free yet improves greatly the practical accuracy of the paired-electron approximation.

\section{Methods}

\begin{figure}[h!]
\centering
\includegraphics[scale=0.4]{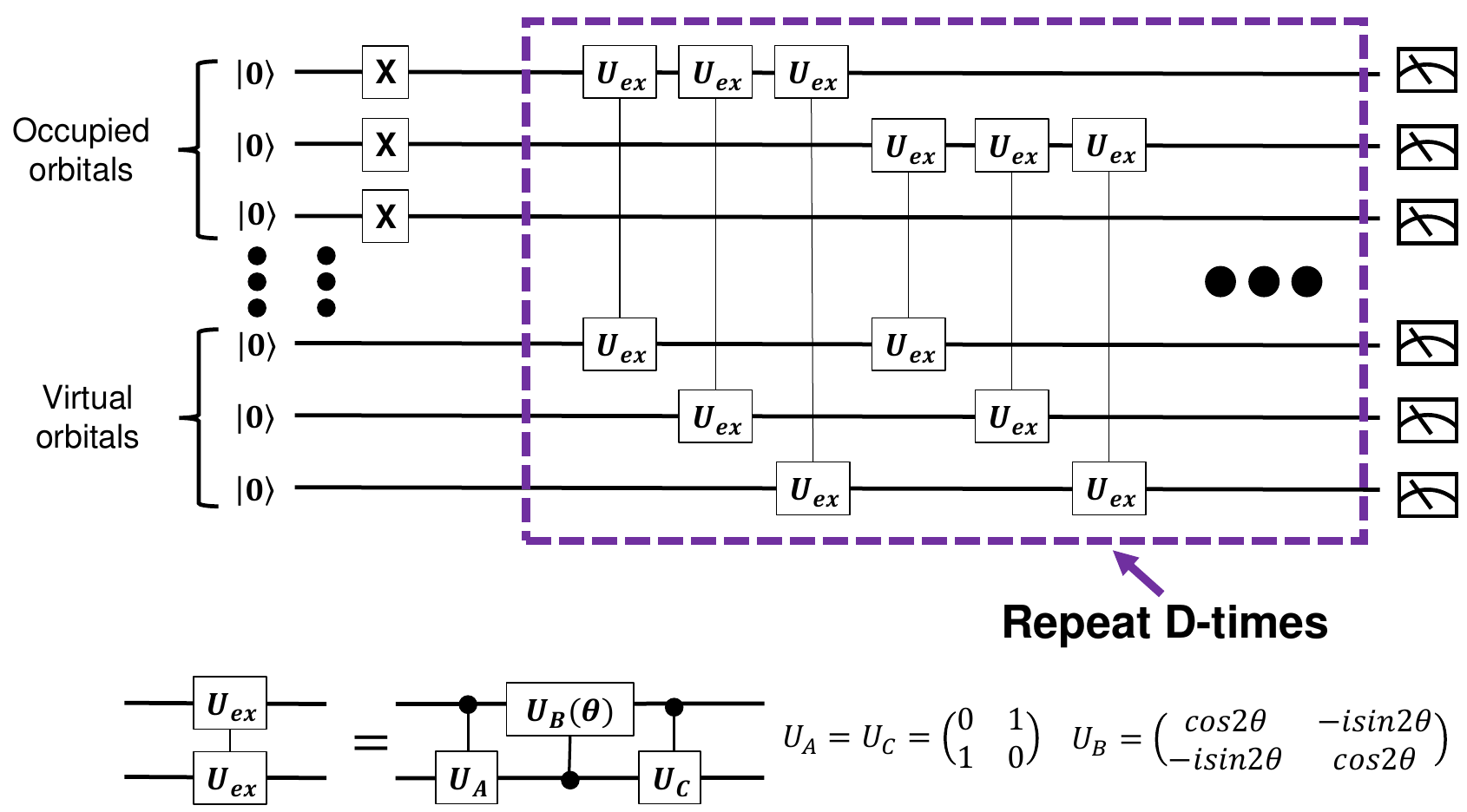}
\caption{Ansatz preparation based on exchange-type gates, as suggested in ref \citeonline{barkoutsos2018quantum}.}
\label{fig:ansatz}
\end{figure}

In our construction, two electrons in one molecular orbital (MO) correspond to one qubit. Accordingly, when $2n$ electrons are contained in $N$ MOs, the qubit ansatz is prepared as
\begin{equation}
\ket{\Psi_0}=\ket{11...1100...00}
\end{equation}
where the first $n$ qubits are in the 1-state while the remaining $N-n$ qubits are in the 0-state.
Then, exchange gates\cite{barkoutsos2018quantum} are applied between all $O$ occupied orbitals ($i,j, \dots$) and $V$ virtual orbitals ($a,b, \dots$) as shown in Fig.\ \ref{fig:ansatz}. The exchange gate between the qubits $i$ and $a$ actually consists of three gates also as shown in the same figure with $i$-to-$a$ CNOT, $a$-to-$i$ controlled $x$-rotation, and $i$-to-$a$ CNOT. Therefore, $OV$ parameters $\left\{ \theta_{ia} \right\}$ assigned for each of the $OV$ exchange gates are used in the ansatz generation, resulting in $OV$ parameters and $3OV$ two-qubit gates in total. This process can be repeated $D$ times with additional sets of parameters with the same structure. The repeating number $D$ can be determined phenomenologically through an optimization process over the entire algorithm. For the molecules tested in this work, $D = 1$ was good enough. While the ansatz preparations based on UCCSD and qubit coupled cluster (QCC) methods have gate complexities of $O(N^3)$ or $O(N^4)$ for a system with $N$ electrons,\cite{Gokhale2019ON3MC} the quantum circuit part in our case can be constructed with only $O(N^2)$ gates and a similarly scaling number of parameters. The efficiency of this exchange-type gates was also confirmed by Barkoutsos et al.\cite{barkoutsos2018quantum}

Now, the final state can be translated into
\begin{align}
\ket{\Psi} =&\prod_{i,a} U_{ia}^{\rm ex}(\theta_{ia})\ket{\Psi_0} =
 c_0\ket{11...1100...00}+c_1\ket{11...1010...00}+ \dots + c_{M}\ket{00...0011...11}
 \label{eq:bosonic_csf}
\end{align}
and the parameters are optimized by adopting the conventional Hamiltonian
\begin{equation}
H = \sum_{pq} \sum_{\sigma} h_{pq} a^{\dagger}_{p\sigma} a_{q\sigma} + \sum_{pqrs} \sum_{\sigma\tau} \frac{1}{2} (ps|qr) a^{\dagger}_{p\sigma} a^{\dagger}_{q\tau} a_{r\tau} a_{s\sigma}
\label{eq:H_original}
\end{equation}
with the one- and two-electron integrals $h_{pq}$ and $(pq|rs)$ in chemists' notation. Here, the spin indices $\sigma$ and $\tau$ supplement the spatial orbital indices $p,q, \dots$ that cover all possible MOs. 
Note that we are only including bosonic pair excitations from $i$ to $a$, which will reduce the Hamiltonian into a simpler form. The details of this reduction can be found elsewhere,\cite{zhao2023} and we will briefly walk through it here for completeness. Because only paired excitations contribute, the summation over the two-electron terms can first be grouped into
\begin{align}
    \sum_{pqrs} =& \sum_{\substack{p=q \\ \ne r=s}}
    + \sum_{\substack{p=r \\ \ne q=s}}
    + \sum_{\substack{p=s \\ \ne q=r}}
    + \sum_{\substack{p=r \\ =q=s}} \\
    =& \sum_{\substack{p=q \\ \ne r=s}}
    + \sum_{\substack{p=r \\ \ne q=s}}
    + \sum_{\substack{p=s, \\ q=r}}
\end{align}
The merge of the last two sums in the first line in the second line is for a later convenience. The first sum is non-vanishing only if $\sigma \ne \tau$, and by introducing $d_{p}^{\dagger}$ and $d_{p}$ as the pair creation and annihilation operators with $d_p = a_{p\beta} a_{p\alpha}$ and $d^\dagger_p = a^\dagger_{p\alpha} a^\dagger_{p\beta}$, we can easily get
\begin{equation}
    \text{(first sum)} = \sum_{p\ne r} (pr|pr) d^\dagger_p d_r
    \label{eq:sum1}
\end{equation}
%
By properly considering commutation relations, we can trivially reach
\begin{align}
    \text{(second sum)} =& 
-\sum_{p\ne q} \sum_{\sigma} \frac{1}{2} (pq|qp) a^{\dagger}_{p\sigma} a_{p\sigma} a^\dagger_{q\sigma} a_{q\sigma} 
= -\sum_{p\ne q} K_{pq} n_p n_q
\label{eq:sum2}
\\
    \text{(third sum)} =& 
\sum_{pq} \sum_{\sigma\tau} \frac{1}{2} (pp|qq) a^{\dagger}_{p\sigma} (a_{p\sigma} a^{\dagger}_{q\tau} - \delta_{pq}\delta_{\sigma\tau}) a_{q\tau}
=\sum_{pq} 2 J_{pq} n_p n_q -\sum_{p} J_{pp} n_p
\label{eq:sum3}
\end{align}
Note that in the second sum, the condition of $p \ne q$ additionally reduced the summation over the spin indices. We also adopted $J_{pq}$ and $K_{pq}$ to respectively represent the Coulomb and the exchange integrals associated with $p$ and $q$ in compact forms, as well as the number operator $n_p=a_{p\alpha}^{\dagger}a_{p\alpha}=a_{p\beta}^{\dagger}a_{p\beta}$. With these, the working expression for the Hamiltonian is obtained as
\begin{equation}
    H=\sum_p (2h_{pp} - J_{pp})n_p + \sum_{p\ne q} K_{pq} d^\dagger_p d_q
    - \sum_{p\ne q} K_{pq} n_p n_q 
    + \sum_{pq} 2J_{pq} n_p n_q
\end{equation}
While we can adopt the set of HF MOs for constructing this Hamiltonian, it will not be an ideal choice for obtaining the molecular energy. Therefore, we performed orbital optimizations toward minimizing $\braket{\Psi| H |\Psi}$ as in ref \citeonline{zhao2023}.

Then, the total energy is calculated as
\begin{align}
E=& \braket{\Psi| H |\Psi} + E_{\mathrm{nB}} 
\end{align}
where $E_{\mathrm{nB}}$ is the energy contributed by the non-bosonic excitation terms with the configurations neglected in Eq.\ (\ref{eq:bosonic_csf}). 

Although the analytic expression of $E_{\mathrm{nB}}$ cannot be derived based on the information available with the pair excitations, its contribution may still be accounted for toward achieving more reliable energy calculations at least at a heuristic level.
We propose to approximate it as
\begin{equation}
E_{\mathrm{nB}} = -\sum_{pqrs}^N\phantom{}^{'} (pr||qs) \left[ \braket{\Psi| a^\dagger_p a_r a^\dagger_r a_p |\Psi}\braket{\Psi| a^\dagger_q a_s a^\dagger_s a_q |\Psi} \right] ^{1/2}
\label{eq:nB_correction}
\end{equation}
where $(pr||qs)$ denotes the conventional electron repulsion integral (ERI), $(pr||qs) = (pr|qs)-(ps|qr)$,\cite{szabo2012modern} related to electron excitations from orbitals ($p,q$) to ($r,s$). The primed sum denotes that a dummy item that will correspond to a paired excitation (namely, with $p=q$ and $r=s$) should be avoided.

This correction was devised with the following reasoning. First, among all missing unpaired configurations, the ones with two or four unpaired electrons (seniority 2 and 4) will contribute most. This is a reasonable assumption when we consider the terms that constitute low order correlation corrections. Also, such configurations can be generated by operating $a^\dagger_r a_p a^\dagger_s a_q$ (with omitted spin indices for brevity) on some doubly occupied configuration. Next, we argue that the energy contribution by an unpaired configuration will be proportional to the associated ERI, $(pr||qs)$. To know the actual contribution, we need its proportionality constant (``amplitude''), but by construction we do not have that information. We finally reason that this unknown amplitude can be approximated by the geometric mean of the two contributions related to $a^\dagger_r a_p$ and $a^\dagger_s a_q$, namely the portions of $\ket{\Psi}$ that have filled $p$ ($q$) orbital and empty $r$ ($s$) orbital. These will be the norms of $a^\dagger_r a_p \ket{\Psi}$ and $a^\dagger_s a_q \ket{\Psi}$, leading to Eq.\ \ref{eq:nB_correction} . After a short algebra, we can also show that Eq.\ \ref{eq:nB_correction}  is equivalent to
\begin{equation}
E_{\mathrm{nB}} = -\sum_{pqrs}^N\phantom{}^{'} (pr||qs) \braket{\Psi| (1-n_r)n_p
  |\Psi}^{1/2}\braket{\Psi| (1-n_s)n_q  |\Psi}^{1/2}
\label{eq:nB_correction2}
\end{equation}
which is useful for the sake of measurements.

In calculating this energy correction with $E_{\rm nB}$, we need a way of fixing the orbital signs. In this work, we aimed to adjust the signs such that the energy becomes as low as possible. In principle, we can test all different combinations of the signs of all orbitals, but doing so will require testing on $\sim 2^N$ combinations with the number of orbitals $N$, which will be unacceptable. Thus, we have taken the following practical tactic. When the number of electron pairs is denoted as $n$, we started by arbitrarily fixing the signs of the $n$-th and the $(n+1)$-th orbitals, which will correspond to the highest occupied and the lowest unoccupied orbitals with the HF picture. Of course, after the further orbital optimizations, HF picture will not last any longer, but the indices inherited from the initial HF calculations still remain. With an index $a$ fixed to $a=(n+1)$, then we walked down the indices over $\{i= (n-1), (n-2), \dots, 1 \}$ together with $i_{+} = i+1$, and at each stage fixed the sign of the $i$-th orbital such that the electron-repulsion integral $(i_+ a | ia)$ becomes positive. The same process was also taken for over $\{a= (n+1), (n+2), \dots, N\}$ by considering $(na_- | na)$ with $a_- =(a-1)$ and by walking up in the index space of a. For these procedures, we can possibly use indexes of canonical orbitals in defining $i$ and $a$. However, doing so does not provide any connection between adjacent orbitals, and sometime $(i_+a | ia)$ or $(na_- | na)$ is too close to zero, which render the sign fixing process rather ill defined. Choosing spatially best overlapping orbital as the neighboring one will be better in this regard, but orbitals are conventionally given with orthogonality intrinsically implemented and using overlap will not be a good tactic in this regard, either. Therefore, we adopted the one-electron matrix $h_{pq}$ to decide the proximity in orbitals. Namely, for any given orbital $i$, we chose $i_+$ such that $i_+$ was not considered before and $h_{ii_+}$ is the maximum. The same can be applied for choosing $a_-$. Apparently, the computational efforts for choosing the sequence of orbitals in this manner scale as $\sim N^2$.

\section{Results}
\begin{figure}[h!]
\centering
\includegraphics[scale=0.3]{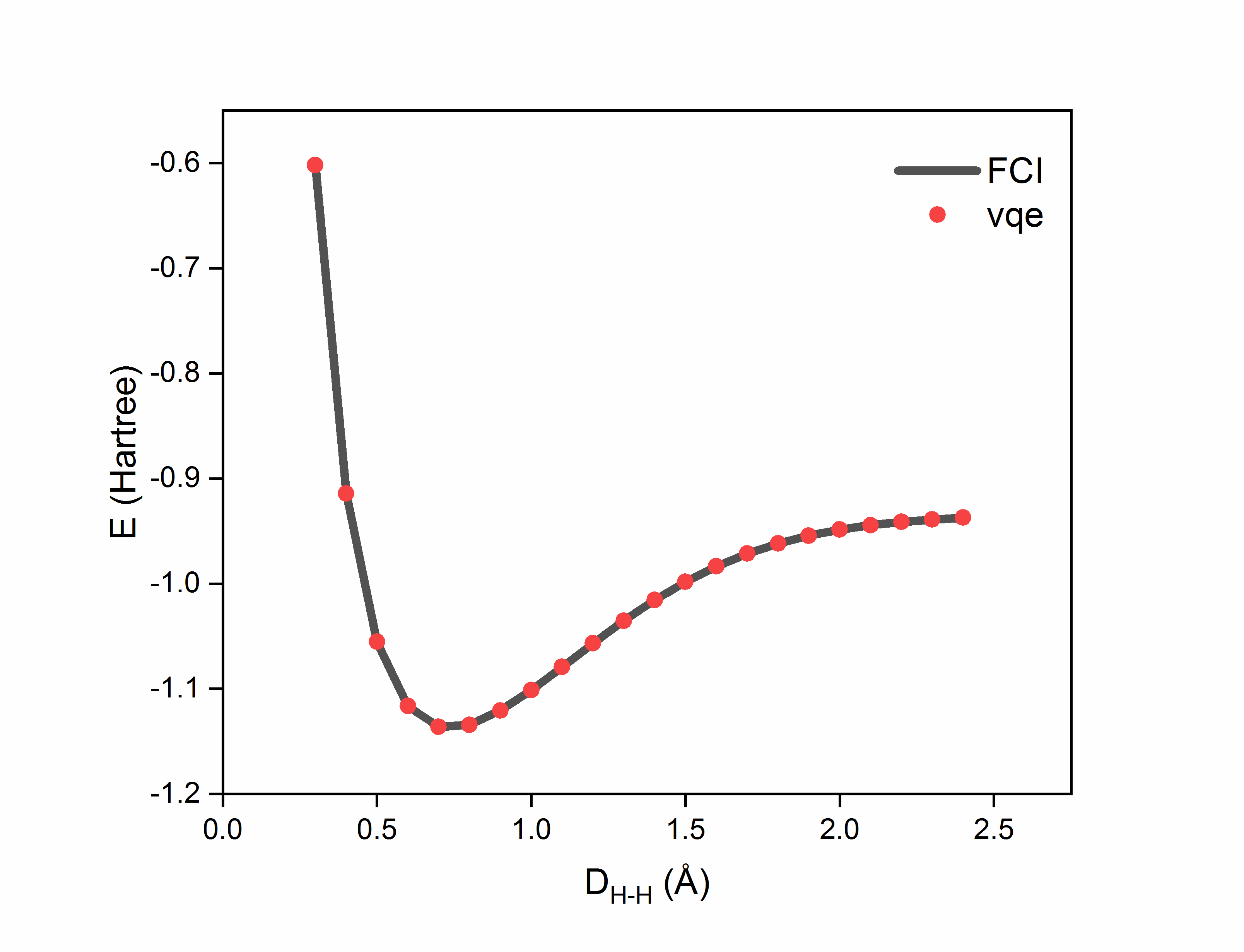}
\caption{Dissociation curve of the H$_2$ molecule. }
\label{fig:h2}
\end{figure}

We first tested our VQE algorithm using H$_2$ with the minimal basis set, STO-3G. Because of the symmetry constraint, this case with only two MOs does not involve any correlations with a non-bosonically excited configuration. Thus, it can serve as a baseline benchmark for our closed-shell quantum simulator algorithm. In terms of the number of qubits, only two were required while four qubits will be needed with the conventional Jordan-Wigner encoding for handling four spin-orbitals. Because there are only one occupied and one virtual orbitals in this case, one exchange gate with one mixing parameter was enough for VQE. The results at varying bond distances, in the range from 0.3 to 2.4 {\AA} is plotted in Figure\ \ref{fig:h2}, together with the corresponding FCI and the HF results. One can clearly see that the VQE result with $\sim$10$^5$ shots is in an excellent agreement with the FCI energies with nearly negligible errors of $\ll$1 milliHartree (mH), evidencing that our VQE algorithm is working properly.

We next tested our scheme by computing the bond dissociation potential curves of a series of small molecules: 
H$_2$O, LiH, N$_2$, and Li$_2$O, with the same STO-3G basis set and the same number of shots for obtaining the expectation values of the Pauli terms. The adopted computational resources for the molecules are listed in Table\ \ref{tab:resources}. In this work, we adopted a quantum simulator IBM-Qiskit,\cite{gadi_aleksandrowicz_2019_2562111} and thus with actual quantum hardware, the optimal numbers of shot averages will differ depending on the fidelity and the coherence time of the hardware. 

\begin{table}
\centering
\begin{tabular}{|p{4cm}||p{2cm}|p{2cm}|p{2cm}|p{2cm}|p{2cm}|}
 \hline

 \hline
 Molecule & H$_2$ & LiH & H$_2$O  & N$_2$ & Li$_2$O \\
 \hline
 No. qubits (occ,vir)   & 2 (1,1)  & 3 (1,2) & 6 (4,2) &  8(5,3)  &  12 (4,8)\\
 No. paramters ($\theta_{ia}$) &   1  &  2 &  8  & 15 & 32 \\
 No. two-qubit-gates &  3  & 6  & 24  & 45 & 96 \\
 \hline

\end{tabular}
\caption{Quantum resource requirements for tested small molecules}
\label{tab:resources}
\end{table}

\begin{figure}[h!]
\centering
\includegraphics[scale=0.3]{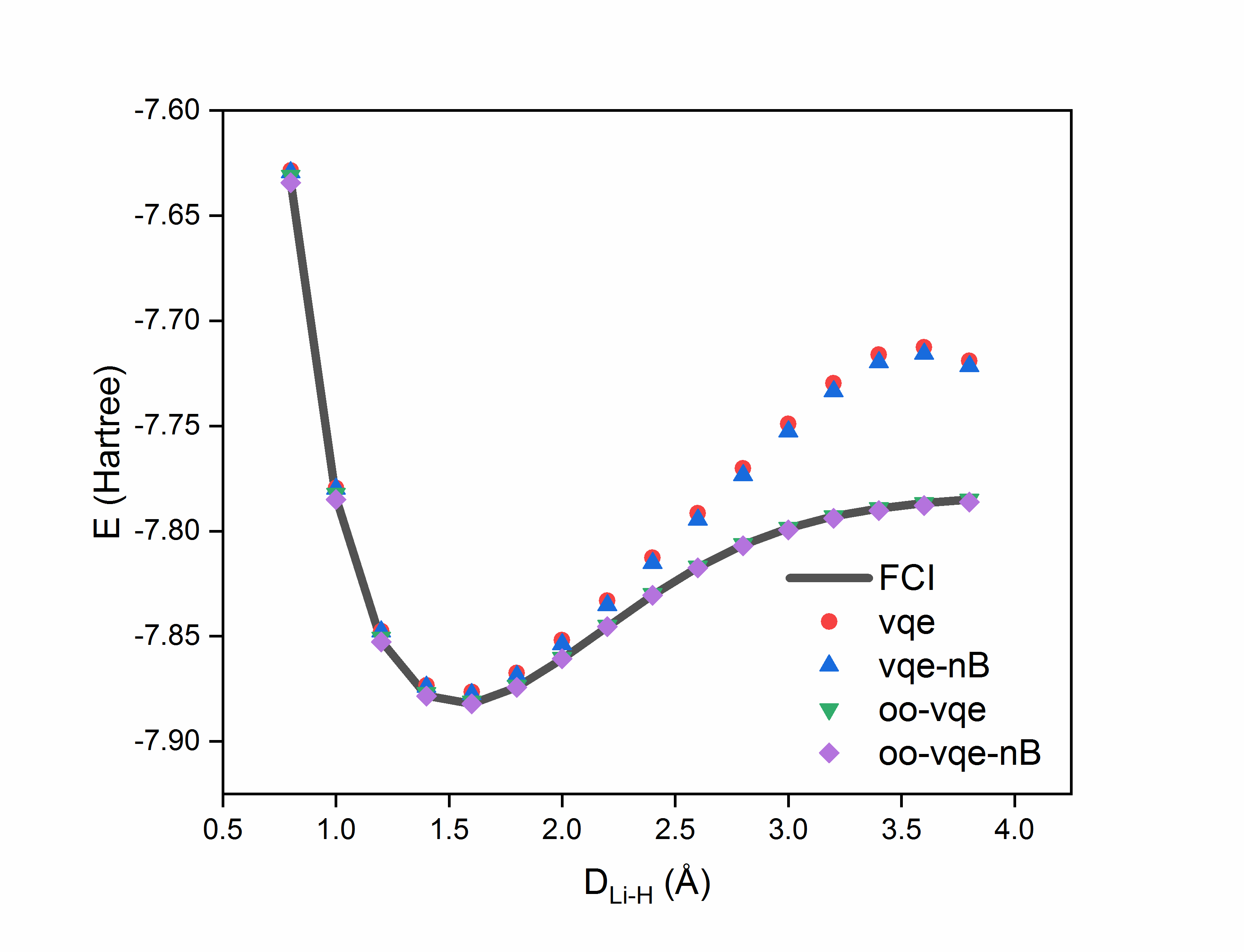}
\caption{LiH ground state energy according to a function of the Li--H distance. }

\label{fig:LiH}
\end{figure}

The first tested molecule, LiH, serves the purpose of checking how much correlation energy can be recovered with our scheme. In this case, there are six MOs with the STO-3G basis. VQE simulations were performed by a circuit consisting of six qubits, corresponding to two occupied and four virtual orbitals. The ground state energies were obtained in the Li--H distance range of 0.5 to 2.4 {\AA} and are shown in Fig.\ \ref{fig:LiH}. As shown in this figure, hereafter, we will designate the VQE energies based on HF orbitals with ``vqe'' and the ones after orbital optimizations with ``oo-vqe''. When $E_{\rm nB}$ in eq \ref{eq:nB_correction2} is added, of course, the energies will be respectively designated as ``vqe-nB'' and ``oo-vqe-nB''. Both vqe and vqe-nB results show good agreements to FCI energy around the distance near the equilibrium separation. However, the errors become quite large at long separations, and orbital optimizations indeed cover this discrepancy rather nicely. This is expected as the spin-restricted HF orbitals should severely fail in such a case, and pair-correlation methods are known to perform well for handling the related nondynamical correlations.\cite{Bartlett2023} In any case, over the entire region, our pair-correlated VQE performs quite well after orbital optimization and $E_{\rm nB}$ does not contribute much with LiH. 

\begin{figure}[h!]
\centering
\includegraphics[scale=0.3]{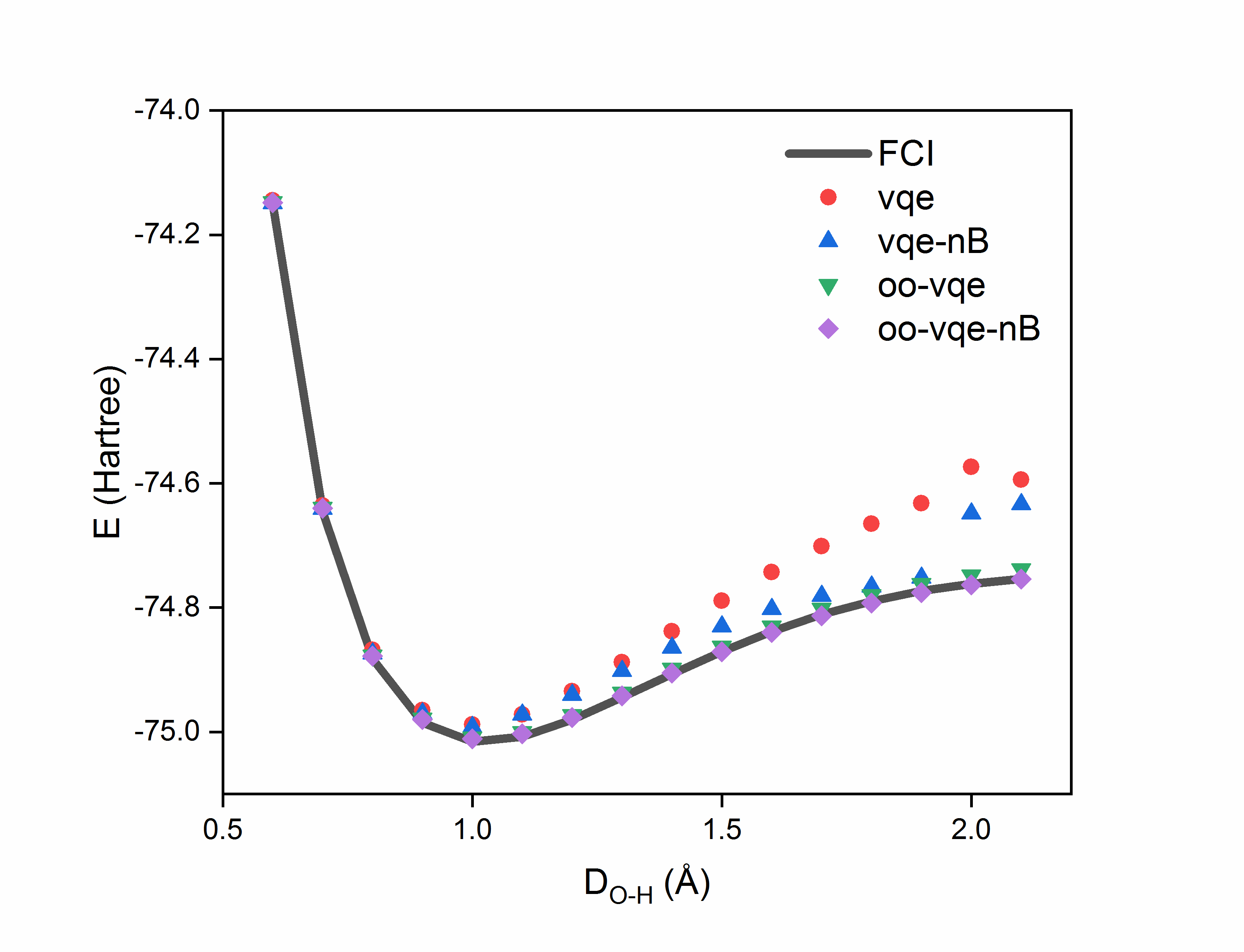}
\caption{H$_2$O ground state energies with differing O--H distances. The two bond lengths were kept identical to each other. }
\label{fig:h2o}
\end{figure}

The next tested molecule, H$_2$O, has seven MOs with the minimal basis set. Of the seven, oxygen $1s$ hardly participates in forming the bonds, and we excluded this core orbital from the VQE calculations via the frozen-core approximation. Hence, VQE was performed by mapping six MOs (four occupied and two virtual ones) to as many qubits. Thus, there were  8 parameters in applying the exchange gates between the occupied and virtual orbitals, with a total of 24 two-qubit gates. Figure\ \ref{fig:h2o} shows how the molecular energy changes by varying the H--O bondlength at a fixed H--O--H angle of 104.45 deg. From this figure, we can again see that the paired ansatz by itself (vqe) displays a significant deviation from the FCI result. Much of the discrepancy is fixed with the subsequent orbital optimization (oo-vqe), and the non-bosonic correction that we propose here almost correctly recovers from the remaining error, with the largest deviation from the FCI curve being barely noticeable from the figure. Interestingly, the non-bosonic correction without performing the orbital optimization (``vqe-nB'') also displays quite a reliable agreement in this case.

\begin{figure}[h!]
\centering
\includegraphics[scale=0.3]{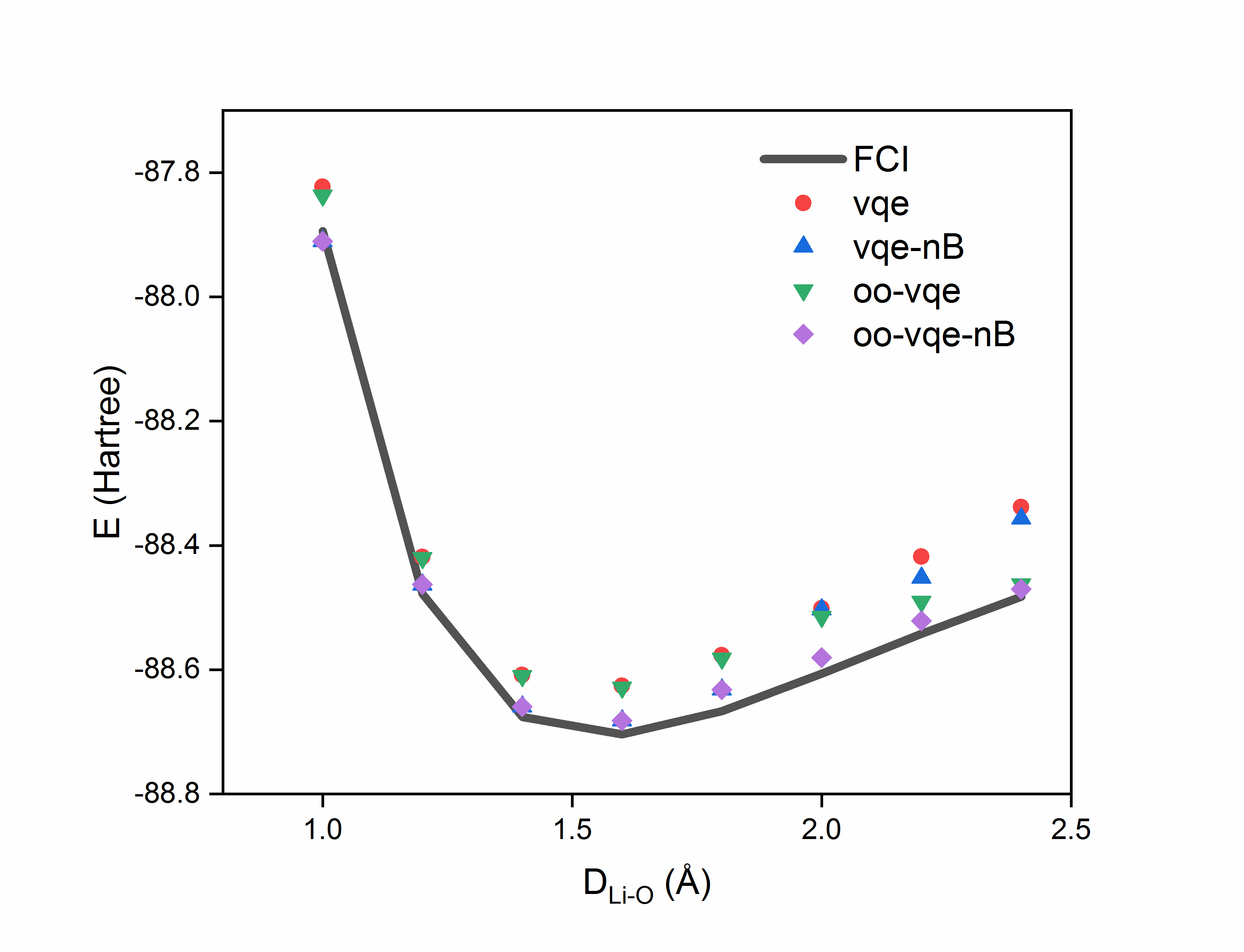}
\caption{Li$_2$O ground state energy at varying Li-O distances.}
\label{fig:Li2O}
\end{figure}

Now, let us move on to a larger system Li$_2$O. Indeed, the molecule is practically related to the operation of Li-air batteries, and enabling quantum simulations of battery materials will be of significant industrial interest.\cite{Rice2021JCP} Similarly to H$_2$O, by applying the frozen core approximation for $1s$ orbitals, we were able to conduct VQE simulations with only 12 qubits. As they represented 4 occupied and 8 virtual orbitals, a total of 32 parameters matched with 96 two-qubit gates were needed. The ground state energies at varying Li--O bond distances are shown in Fig.\ \ref{fig:Li2O}. Compared to H$_2$O, because there are more virtual orbitals available, we can expect that the contribution of the non-bosonic excitations will be larger in this case. Indeed, from the figure, we can see that vqe or oo-vqe does not reproduce the FCI PES that well. Instead, oo-vqe-nB is reproducing FCI much better with errors within $\sim$10 mH. How well each electronic configuration in the CI expansion is represented in VQE by the non-bosonic correction terms will be discussed in more details in a later section.

\begin{figure}[h!]
\centering
\includegraphics[scale=0.3]{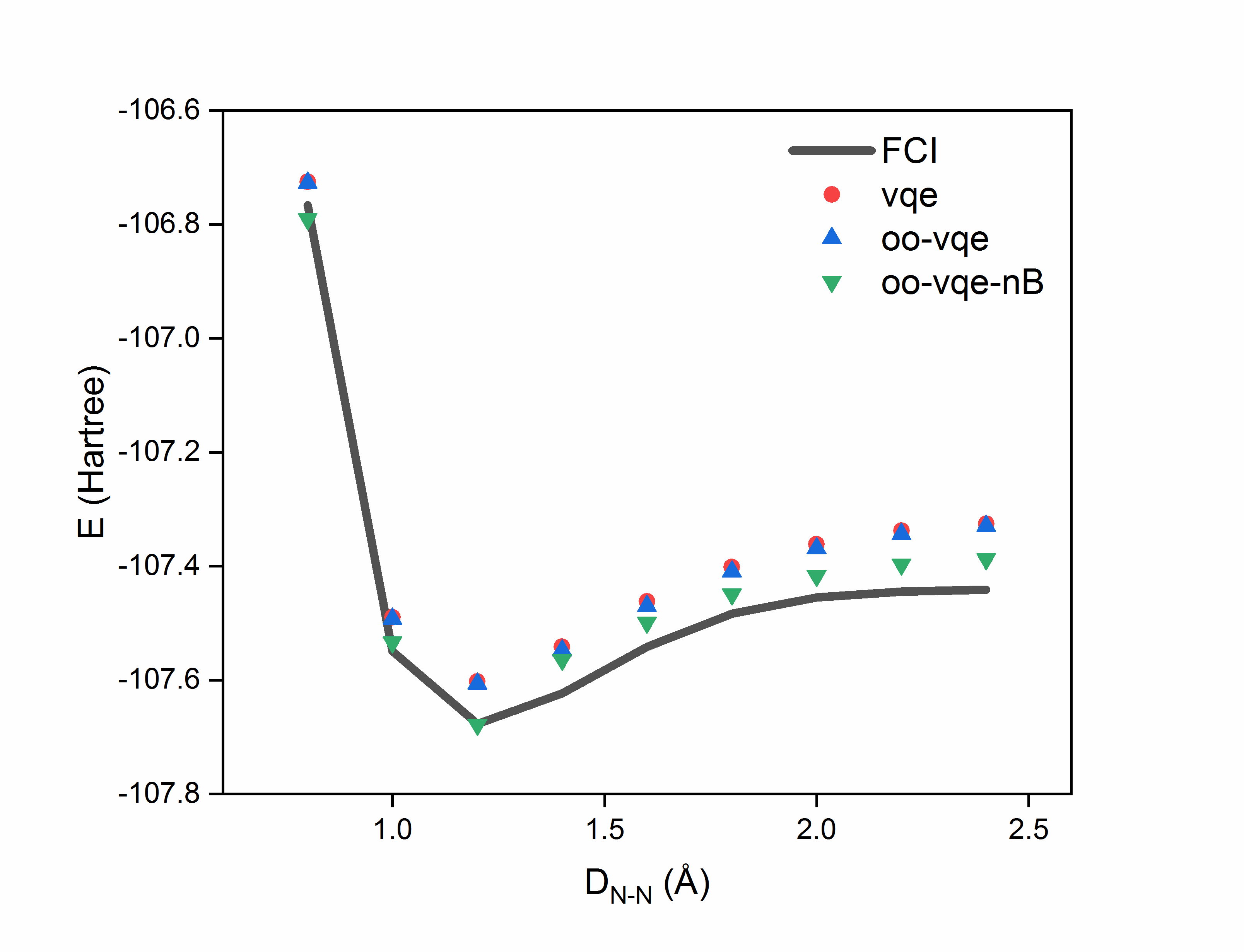}
\caption{Ground state energies of N$_2$ at varying bond lengths. }
\label{fig:n2}
\end{figure}

Now, let us consider the dissociation curve of N$_2$ to further confirm the utility of our approach. In fact, N$_2$ with its triple bond has been considered as one of the most difficult systems to model with electronic structure theories, and accordingly, it will act as a stringent test case for us. Not surprisingly, even the CCSD(T) method fails drastically in describing this triple bond dissociation because it is still a single-reference approach (Fig.\ \ref{fig:n2}). On the contrary, our VQE results with the frozen core approximation reasonably reproduce the FCI results, with the equilibrium bond length and the entire energetic features being in good agreements. We note, however, that the energies stretched from the equilibrium geometry ($R_{\mathrm{N-N}} \sim$ 1.2 \AA) displays an error of 30 -- 50 mH. We will defer commenting on this quantitative discrepancy to a later section, as a more detailed analysis about this mismatch will be covered in the Discussion section.   

\section{Discussion and Summary}

In order to see how our method that separately treats bosonic and non-bosonic excitations constructs the electronic structure, with Li$_2$O at 1.6 {\AA} separation, we compared the populations of each excited configuration with FCI and VQE. The largest population among bosonically excited configurations was associated with the excitation involving the 7-th and the 13-th MOs (in the order of canonical orbital energies), with the FCI population of 0.11156. The same population from vqe can be compared as $\sim$0.11244, which is in close agreement with the FCI value. The second largest population, involving an excitation between the 6-th and the 12-th HF MOs, has essentially the same population with 0.11156 for FCI and 0.11142 for vqe. The third populations similarly compare favorably as 0.04982 and 0.04628. Therefore, FCI and our VQE scheme show very similar electronic structures.

The situation with N$_2$ was somewhat different. At its minimum energy geometry ($\sim$1.2 \AA), the contribution by a quadruple excitation term was quite important with FCI. Namely, a double pair excitation from the 6-th and the 7-th (occuied) MOs to the 8-th and the 9-th (virtual) MOs contributed by $\sim$30\% as large as the most important bosonic double excitation (7-th $\rightarrow$ 9-th). While our method could still accommodate such an excitation in pairs and thus the result was not too bad, the situation will likely increase the importance of pair-broken multiple excitations. Thus, our approach started to deviate from the exact answer in stretched geometries as higher excitations become more important. Of course, the fidelity of the simple correction with Eq \ref{eq:nB_correction} will also deviate. Thus, we note that treating a triple bond still remains as a difficult problem.

We also wish to point out the effects and the limitations of our bosonic mapping and non-bosonic correction. It has been discussed that the bosonic mapping is a powerful tool and can qualitatively reproduce the dissociation curves of simple molecules.\cite{elfving2020simulating} However, it is destined to underestimate the correlation energy because it only includes a subset of properly treated excitations. Interestingly, our non-bosonic correction mostly shows quite good agreements with exact answers at least for the single-bonded molecules, as exemplified with H$_2$O and Li$_2$O. In contrast, in the cases of molecules involving double or triple bonds, our correction term did not work that well. Especially, in the case of N$_2$ with a triple bond (Fig.\ \ref{fig:n2}), similar results were obtained whether the correction term was added or not. Even in such cases, however, our approach can still be considered meaningful in terms of reducing the required resources for simulations. Although several reports have shown that quantum unitary coupled cluster singles and doubles (UCCSD) can reproduce the dissociation curves of some small single-bonded molecules and even N$_2$ within a few mH error,\cite{sokolov2020quantum, gomez2016recoupling, lee2018generalized, mizukami2019orbital} the required number of gates were about $10^4$ -- $10^5$,\cite{sokolov2020quantum} which are still distant from the practical applicability with the currently available NISQ devices. It is also interesting to note that various symmetry-restricted versions of quantum UCCSD approaches show errors in the tens of mH regime.\cite{sokolov2020quantum} Our model actually provided curves with an at least similar level of errors, but with only the half number of qubits and tens of two-qubit gates. Indeed, our resource requirements will likely be much more reasonable with the presently available devices. 

In summary, we have proposed a VQE quantum simulation method that uses a qubit to map a spatial MO in closed-shell molecules with a rather simple correction. The method requires only a half number of qubits compared with conventional Jordan-Wigner or Bravyi-Kitaev mapping based methods. Using the method, dissociation curves of LiH, H$_2$O, and Li$_2$O were obtained, demonstrating errors within $\sim$10 mH in comparison with FCI. The number of gates and optimization parameters are proportional to $~O(N^2)$, which is significantly less than the more conventional $~O(N^3)$ or $~O(N^4)$ scaling\cite{Gokhale2019ON3MC} and is reasonably accessible with actually available physical qubit systems. We wish that our method can be of help in further advancing techniques in the NISQ era that we have already entered.

\bibliography{sample}

\section*{Acknowledgments}

This work was supported by Hyundai Motor Company (Hyundai-KAIST Joint Quantum Algorithm Research Project). We thank Dr. Seung Hyun Hong and Dr. Jongkook Lee at Hyundai Motor Company for enlightening discussions. We also thank Dr. Luning Zhao and Dr. Joshua Goings at IonQ for useful discussions.

\section*{Competing Interests}

The authors declare no competing interests.

\end{document}